\documentclass[11pt,a4paper]{article}
\pdfoutput=1
\usepackage{jinstpub}

\usepackage{xspace} 
\usepackage[usenames,dvipsnames]{color}
\usepackage{booktabs,graphicx,mathrsfs,verbatim,amsmath,units,soul}
\usepackage{ulem}
\usepackage{xspace} 
\usepackage{amsmath} 
\usepackage{upgreek}
\usepackage{gensymb}
\usepackage{tabularx}
\usepackage{fullwidth}
\usepackage{multirow}
\usepackage [autostyle, english = american]{csquotes}
\usepackage{subcaption}
\usepackage{graphicx}
\usepackage{xcolor}
\MakeOuterQuote{"}

\newcolumntype{Y}{>{\centering\arraybackslash}X}

\newcommand\figref[1]{Figure\,\ref{#1}}
\newcommand\eqnref[1]{Eq.\,\ref{#1}}
\newcommand\tabref[1]{Table\,\ref{#1}}
\newcommand\secref[1]{Section\,\ref{#1}}

\newcommand\citeref[1]{Ref.\,\cite{#1}}

\newcommand{\RNum}[1]{\uppercase\expandafter{\romannumeral #1\relax}}

\begin{document}

\title{Characterization of photomultiplier tubes with a realistic model through GPU-boosted simulation}



\author[a, 1]{M.~Anthony%
\note{Corresponding author.},}
\author[a]{E.~Aprile,}
\author[b]{L.~Grandi,}
\author[a]{Q.~Lin,}
\author[b, c]{R.~Saldanha}

\affiliation[a]{Physics Department, Columbia University, New York, NY 10027, USA}
\affiliation[b]{Department of Physics \& Kavli Institute for Cosmological Physics, University of Chicago, Chicago, IL 60637, USA}
\affiliation[c]{Pacific Northwest National Laboratory Richland, WA, 99354, USA}


\emailAdd{mda2149@columbia.edu}

\abstract{
The accurate characterization of a photomultiplier tube (PMT) is crucial in a wide-variety of applications.  However, current methods do not give fully accurate representations of the response of a PMT, especially at very low light levels.  In this work, we present a new and more realistic model of the response of a PMT, called the cascade model, and use it to characterize two different PMTs at various voltages and light levels.  The cascade model is shown to outperform the more common Gaussian model in almost all circumstances and to agree well with a newly introduced model independent approach.  The technical and computational challenges of this model are also presented along with the employed solution of developing a robust GPU-based analysis framework for this and other non-analytical models.
}

\maketitle


\section{\label{sec:appendix}Introduction}

Photomultiplier tubes (PMTs) are widely used to detect low levels of light in many fields of physics.  However, despite their ubiquitousness, calibration and characterization of the single photoelectron (SPE) charge response of PMTs remains in a fairly basic state.  PMTs are very complicated devices yet they are often treated with a simple approximation: that the SPE charge response is Gaussian \cite{bellamy, dossi}. While this approximation is satisfactory for specific PMTs within certain voltage ranges, it is far from true in general.  Since the Gaussian distribution is not bounded below by zero, the response function cannot be correct for a SPE and oftentimes, when PMT calibrations are performed with low PMT voltages, the response function will have a large probability of producing a non-physical signal. 

Several alternatives have been proposed to improve upon existing methods for determining the single photoelectron response.  An empirical approach is presented in \citeref{haas} but is only relevant when the height of the PMT output is needed and not the integral of the pulse.  Another model independent approach is presented in \citeref{saldanha}.  The model independent approach provides a simple way to accurately determine the mean and variance of the single photoelectron response function.  In many cases, the mean and variance of the SPE response are  enough since at moderate numbers of photoelectrons the response function converges to a Gaussian described by these parameters.  However, at small numbers of photoelectrons, it is important to account completely for the SPE response shape.  Additionally, the results of the model independent method become more susceptible to bias when the background distribution width is large and the PMT gain is low and it requires a consistent and dedicated background measurement, which is not always possible.  Background, in this work, is used to describe all signals that are not induced by the laser or diode, such as noise from the electronics, dark counts, or photoelectrons from light sources other than the laser or diode.  Additionally, several non-analytical models have been proposed for the single photoelectron response of a PMT that account for the details of electron multiplication via the dynode structure \cite{lombard,prescott}.  However, given the difficulty of performing parameter estimation without a analytical description of the model, these models are typically not used for measurements of individual photomultiplier responses.

In this work, we propose a more realistic model, henceforth referred to as the cascade model, which aims at capturing the actual behavior and mechanics of the PMT.  The model does not have an analytical form that can be used for parameter estimation but rather relies on running a Monte Carlo (MC) simulation with each set of parameters under test to find the posterior and the best-fit parameters given the data.  These MC simulations come with a high computational cost since they are run at each iteration of the fit so it was necessary to build the analysis framework on a GPU based server rather than the traditional CPU based server.

\section{\label{sec:level2}Method}

\subsection{\label{sec:level2-4}The Cascade Single Photoelectron Charge Response Model}

With almost countless varieties of PMTs used in different settings, it is impossible to describe a single model that will accurately characterize all PMTs under all circumstances.  However, in this paper, a SPE response model is presented that has been found to be successful for two very different PMTs and which is physically motivated according to \citeref{hammamatsu}.  

\begin{figure}[h]
\centering
\includegraphics[width=7cm]{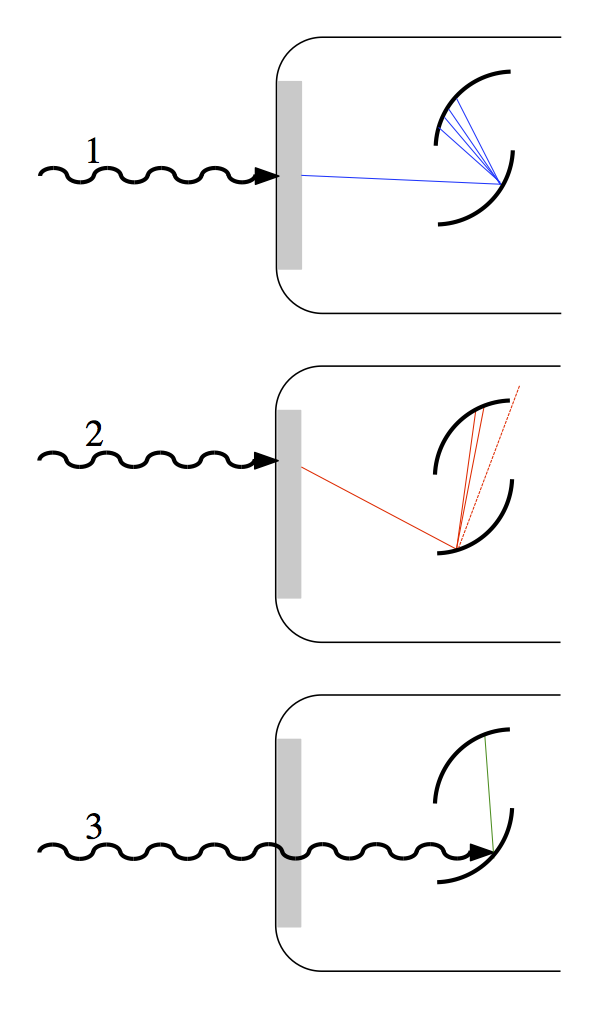}
\caption{The three possible scenarios for photoelectrons in the cascade model.  Scenario 1 shows the standard full-amplification: a photon is absorbed in the photocathode and an electron is amplified through the dynode chain.  Scenario 2 shows a non-ideal trajectory: a photon is absorbed in the photocathode but the electron follows a slightly different trajectory, due to field imperfections, and suffers a slightly lower amplification.  Scenario 3 shows a photon passing through the photocathode and releasing an electron on the first dynode.  Note that in this scenario  the amplification at each dynode may depend on where the incident photon strikes the first dynode.}
\label{fig:fig-pmt_diagram}
\end{figure}

In the cascade model, there are three different physical processes that can produce an output signal.  Each of these scenarios is depicted in \figref{fig:fig-pmt_diagram}.
\begin{enumerate}
	\item Full amplification: this is the most common process for producing a signal from the PMT.  This occurs when a photon is absorbed by the photocathode which then releases an electron (referred to as a photoelectron).  This electron is then accelerated to the first of the multiple dynodes found inside of the PMT.  This electron will then strike the surface of the dynode and release more electrons in the process.  These secondary electrons are then accelerated towards the second dynode.  This process continues through all the dynode stages and results in a signal that is proportional to the number of photons initially absorbed by the photocathode.
    \item Bad trajectory amplification of photoelectrons from the photocathode: this is very similar to full amplification with a single important change.  The electron released from the photocathode may follow a non-ideal trajectory which will result in secondary electrons potentially not reaching the next stage of amplification.  This will ultimately result in lower amplification and is caused by electric field imperfections in the PMT.  
    \item Amplification from direct excitation of the first dynode: this occurs when a photon passes through the photocathode and strikes the first dynode, in turn releasing an electron.  This electron then follows the chain of amplification, albeit with one less dynode.  The initial electron may also follow a non-ideal trajectory which results in smaller than normal amplification even accounting for the loss of a dynode stage.
\end{enumerate}

To approximate these three physical processes in the SPE response, eight parameters were used:

\begin{itemize}
    \item $p_{pc}$ (``photocathode''): the probability that an incident photon produces a photoelectron from the photocathode that is amplified.
    \item $p_{fd}$ (``first dynode''): the probability that an incident photon produces a photoelectron from the first dynode that is amplified.  Note that an incident photon cannot create a photoelectron on both the photocathode and the first dynode.
    \item $p_{bt}$ (``bad trajectory''): the probability that a photoelectron from the photocathode will follow a non-ideal trajectory through the dynodes and will require a correction to the resulting amplification.
    \item $\mu_{epd}$, $\sigma_{epd}^2$ (``electrons per dynode''): the mean and variance of the truncated discrete Gaussian, used to find how many secondary electrons are produced by each incoming electron at each dynode stage, for the smallest electric field in the dynode chain.  These parameters are increased linearly with the electric field at each dynode stage.
    \item $p_c$ (``continue''): the probability that secondary electrons escape the surface of the dynode and reach the following dynode.
    \item $c_{fd}$, $c_{bt}$: the corrections applied to $p_c$ accounting for differences in photoelectron amplification from the first dynode and for underamplification due to bad trajectories.

\end{itemize}






The photoelectron of the SPE response in the cascade model has two potential points of origin: (1) the photocathode or (2) the first dynode.  This implies that the origination is described by a binomial process ($B(n, p)$) with a single trial.

\begin{equation}
n_{pc} \sim B\left(n=1; p=\frac{p_{pc}}{p_{pc}+p_{fd}}\right), \qquad
n_{fd} = 1 - n_{pc}.
\end{equation}

In the above equation, $n_{pc}$ accounts for all electrons coming from the photocathode and $n_{fd}$ accounts for all electrons coming directly from the first dynode.


It is important to note that certain PMTs are found to produce two photoelectrons instead of a single photoelectron at the photocathode with a measured probability, $p_{DPE}$, at certain wavelengths of incident light.  A measurement of this effect is described in \citeref{dpe}.  One can simply account for this double photoelectron effect by adding a binomial process.

\begin{equation}
n_{pc} \leftarrow n_{pc} + B(n=n_{pc}, p=p_{DPE}).
\end{equation}

Further dividing electrons from the photocathode, the model assumes a fixed probability that the electron will follow a bad trajectory.


\begin{equation}
n_{bt} \sim B(n=n_{pc}, p=p_{bt}), \qquad
n_{fa} = n_{pc} - n_{bt}.
\end{equation}

In the above equation, $n_{bt}$ is the number of electrons from the photocathode that follow a bad trajectory, resulting in underamplification, and $n_{fa}$ is the number of electrons that are fully amplified from the photocathode through the entire dynode chain.


With all three potential signal sources accounted for, one can now consider the dynode chain.  For the dynode chain, it is assumed that the electrons follow a Galton-Watson branching process as described in \citeref{tan}.  However, instead of the Poisson distribution as described in \citeref{tan}, the model assumes that the the number of secondary electrons at each dynode stage is described by a discrete Gaussian ($DG(\mu,\sigma^2)$), as described in \citeref{dwarakanath}, and a binomial process (with probability of success $p_c$) to be as general as possible (since the shape and variance of a Poisson distribution are fixed by its mean).  This iterative process is described in \eqnref{gw_1} and \eqnref{gw_2}.  In these equations, $h_{i}$ is the number of secondary electrons leaving the $i^{th}$ dynode while $m_{i}$ is the number of electrons that reach the $i^{th}$ dynode.

\begin{equation}
\label{gw_1}
h_{i} \sim DG(\mu=m_{i} \mu_{epd}, \sigma^2 = m_{i} \sigma^2_{epd}).
\end{equation}
\begin{equation}
\label{gw_2}
m_{i+1} \sim B(n=h_{i}, p=p_{c}).
\end{equation}

For the bad trajectory electrons from the photocathode and the electrons from the direct excitation of the first dynode, the Galton-Watson process is modified such that $p_{c} \rightarrow p_{c} c_{fd}$ or $p_{c} \rightarrow p_{c} c_{bt}$ to account for their non-ideal trajectory.  This correction is applied identically to each dynode in the chain.  Differences in the electric fields between dynodes are accounted for by proportionally increasing the mean and variance of the discrete Gaussian ($\mu_{epd}$ and $\sigma^2_{epd}$ represent the mean and variance of the Galton-Watson process for the smallest electric field in the chain).

While there is not an analytical function to describe the SPE response in the cascade model, we can approximate the probability distribution function (PDF) of the SPE response via MC simulations.

\subsection{\label{sec:level2-1}Statistical Treatment}

To perform parameter estimation, one needs to be able to define a likelihood function.  Since the cascade model is not analytical, one is forced to use a binned likelihood with a modification: instead of using the integral of the PDF to determine the expected number of events in a bin given the parameters under test, one runs a MC to approximate this expectation. 

\begin{equation}
\label{bin_estimate}
\hat{b}_i = \frac{M \cdot N_i}{N}.
\end{equation}

In \eqnref{bin_estimate}, $\hat{b}_i$ is the estimated expected number of events in the given bin for an experiment with $M$ total data points given the current model with the current parameters under test, $N_i$ is number of MC events under those conditions that fell into bin $i$, and $N$ is the total number of MC events run.

With our binned data, $b_i$, and with an approximation for the expectation of each bin, $\hat{b}_i$, one can approximate the log-likelihood:

\begin{equation} \label{eq:log-likelihood}
ln(\mathcal{L}) = \sum_i ln(\mathcal{L}_i) = \sum_i ( b_i ln(\hat{b}_i) - \hat{b}_i - ln(b_i!)).
\end{equation}

This log-likelihood can be modified to account for priors on certain fit parameters.  For example, if one has an independent measurement of the background, one could analyze this dataset separately and then use a prior to constrain $\mu_{bkg}$ and $\sigma_{bkg}$.

Notice that this treatment has three potential drawbacks.  First, $\hat{b}_i$ is a random variable with a variance given by the following equation where $p_i$ is the true probability that a given event falls in bin $i$.

\begin{equation}
\sigma_{\hat{b}_i}^2 = \sigma_{\hat{p}_i}^2 M^2 = \frac{p_i M^2}{N}.
\end{equation}

This implies that the log-likelihood is also a random variable.  In other words, for the same parameters under test one will get different values for the log-likelihood.  It is important to be aware of this effect when using this method and to ensure that the fluctuations in log-likelihood are small.  It is also important to note that these fluctuations will not affect the end result but could pose technical challenges for given choices of minimizers (particularly ones that are dependent on the gradient of the log-likelihood) and slow down convergence.  It is recommended to use minimizers based on genetic algorithms or a Markov Chain Monte Carlo (MCMC) to perform the parameter estimation as gradient-based minimizers will perform poorly when the fluctuations are not negligible.

Two simple solutions to reduce the fluctuations in log-likelihood and improve the convergence speed are to increase the number of Monte Carlo iterations or to increase the size of each bin.  A less desirable but alternate solution when performing a Bayesian analysis is to suppress the log-likelihood at each stage of the iteration, artificially decreasing the fluctuations.  This likelihood suppression allows for faster convergence and is useful if it is unreasonable to increase MC statistics but it does imply that one is no longer sampling from the posterior but from a widened version of it.

Second, if $p_i$ for a given bin is nearly zero it is possible that the Monte Carlo run will produce no events in that bin.  This implies that $\hat{p}_i$ and $\hat{b}_i$ would be exactly equal to zero which is unacceptable for the Poisson distribution implicit in the binned log-likelihood.  Again, there are many ways to handle this type of issue but the simplest is to alter the binning such that each $p_i$ is approximately the same. 

The third drawback is that the computational cost of running a large MC on each iteration of a fit is extremely high.  This issue and the solution used are discussed in more detail in \secref{sec:level2-2}.

To perform the parameter estimation in this paper, the package emcee was used \cite{dfm}.  Each log-likelihood call was made using eight million MC trials to form the histogram needed for \eqnref{eq:log-likelihood} to ensure that log-likelihood fluctuations were on the order of $\frac{1}{2}$.


\subsection{\label{sec:level2-2}Technical Feasibility with High-Performance Graphical Processing Units}

While using Monte Carlo simulations to approximate the PDF of a model allows for flexibility in the model used, it comes with a very large computational price.  Using a high statistic (greater than 1M events) MC on each iteration of a fit will be extremely slow on a standard CPU server, even accounting for the possibility of parallelization.

However, graphical processing units (GPUs) are ideally suited for running MC simulations where each MC trial is independent of the others.  This is because GPUs contain thousands of cores that can be used simultaneously.  Even though each GPU core in itself is less powerful than a standard CPU core, it is still typical to see speed increases on the order of 100--1000x simply due to the parallelization of the GPU.  The realized speed increase depends on the MC simulation itself and the hardware used.

Development of GPUs has expanded rapidly over the last decade, especially for scientific computing.  There is now affordable hardware from multiple companies, several libraries for easy development, and cloud-based GPU services.

For this analysis, a custom GPU based server was designed and built.  This server includes six NVIDIA GTX 1080 cards which gives the server a top speed of roughly 54 TFLOPS.  CPUs typically fall in the range of tens to hundreds of GFLOPS.  The MC that is run in each log-likelihood call is written in CUDA C.  The time required for each fit was roughly one hour on a single GPU card for the implementation used.  By using a single GPU card per fit, six analyses could be performed in parallel.

\subsection{\label{sec:level2-3}Data Collection}

Low light level data was used from two independent experiments using two different methods of data collection and PMTs.  The first set of data was provided by the experiment described in \citeref{saldanha}.  This data is from a Hamamatsu R11410, a 3 inch PMT, the low-background version of which was used in the XENON1T experiment \cite{xenon1t}. The PMT was operated in a dark box with a 405 nm pulsed laser behind a filter with an attenuation factor $\eta$.  By changing $\eta$, one can change the mean number of incident photons.  Background measurements were also taken for this data in the exact operating conditions except with the laser light blocked.

The second set of data is from the neriX detector described in \citeref{goetzke}.  This data is from a Hamamatsu R6041-406 SEL 2 inch PMT in LXe illuminated by a blue pulsed LED located inside the detector.  The PMT used to collect this data operates at a significantly lower gain than the PMT used in \citeref{saldanha} and has worse noise conditions.  Also, identical conditions during background measurements could not be guaranteed and therefore the model independent approach could not be used to characterize this PMT.

In both experiments, the digitized waveforms were integrated with consistent acquisition windows.

Also, the light used to illuminate the PMTs in both experiments had a wavelength larger than 400 nm so double photoelectron (DPE) effects were not included \cite{dpe}.  As mentioned in \secref{sec:level2-4}, DPE effects can straight-forwardly be added to the cascade model if needed.

\section{\label{sec:level3}Results}

\subsection{\label{sec:levelres-char}Response Characterizations}

In this work, three methods were used to characterize the PMTs for which data was collected.  The first was the cascade model, for which the SPE response was described in detail in \secref{sec:level2-4}.  With the model of the SPE response, we can approximate the response of larger signals by convolving the SPE response function ($f_1$ in \eqnref{spe_convolution}) with itself for the number of photoelectrons needed.  Finally, one must consider detector specific effects by convolving the signal with the background spectrum ($f_0$ as defined in \eqnref{bkg_spec}) which we will assume follows a Gaussian distribution ($N(\mu, \sigma^2)$).  In this work, the background is approximated as Gaussian from independent measurements.

\begin{equation}
\label{bkg_spec}
f_0(x) = N(\mu=\mu_{bkg}, \sigma^2=\sigma^2_{bkg}).
\end{equation}
\begin{equation}
\label{spe_convolution}
f_n(x) = f_0(x) \circledast \overbrace{f_1(x) \circledast f_1(x) \circledast \ldots \circledast f_1(x)}^{\text{n times}}.
\end{equation}

To perform parameter estimation, one must consider how the PMT is illuminated.  Since low light levels are used, one expects the number of photoelectrons produced per light pulse to follow a Poisson distribution with a mean $\lambda$ ($P(k, \mu)$).  One then combines the individual contributions to define the PDF of the full spectrum at a certain light level that will be used (\eqnref{cascade_fit_full}).

\begin{equation}
\label{cascade_fit_full}
f(x) = P(k=0, \mu=\lambda) \cdot f_0(x) + \sum^{\infty}_{i=1} (P(k=i, \mu=\lambda) \cdot f_i(x)).
\end{equation}

The second method used was the model independent characterization, which is described in detail in \citeref{saldanha}.  The model independent method uses the statistical properties of the laser calibration charge spectra and a background-only charge spectra to estimate the mean and variance of the single photoelectron response, as well as the mean number of photoelectrons produced per light pulse. This method has the advantage that it does not assume any specific functional form for the SPE response as the PMT response converges to a Gaussian for signals with more than roughly five to ten photoelectrons. However the method requires a dedicated background measurement in identical operating conditions, and additional parameters may be needed if one needs to simulate the full functional response of a PMT at very low light levels.

The third method used to characterize the PMTs was the Gaussian approximation with an underamplified peak.  In this case, the SPE response is the sum of two  Gaussians - one representing fully-amplified photoelectrons and the other representing underamplified photoelectrons.  One can estimate larger signals in the same way as the cascade model: by convolving the SPE response function ($s_1$ as defined in \eqnref{gaussian_spe}) with itself for the number of photoelectrons needed and then with the background as shown in \eqnref{gaussian_mpe}.

\begin{equation}
\label{gaussian_spe}
s_1(x) = N(\mu=\mu_1, \sigma^2=\sigma^2_1) + w \cdot N(\mu=\mu_u, \sigma^2=\sigma^2_u), \, 0 \leq w \leq 1.
\end{equation}
\begin{equation}
\label{gaussian_mpe}
g_n(x) = f_0(x) \circledast \overbrace{s_1(x) \circledast s_1(x) \circledast \ldots \circledast s_1(x)}^{\text{n times}}.
\end{equation}

Finally, to produce the PDF for parameter estimation, one defines a mean number of photoelectrons per pulse and sums each peaks individual contributions weighted by a Poisson distribution (\eqnref{gaussian_pdf}).  This is done in the same way as the cascade model.

\begin{equation}
\label{gaussian_pdf}
g(x) = P(k=0, \mu=\lambda) \cdot f_0(x) + \sum^{\infty}_{i=1} (P(k=i, \mu=\lambda) \cdot g_i(x)).
\end{equation}

In these equations, $\lambda$ is the mean number of PE and $\mu_u$ and $\sigma_u$ are the mean and standard deviation of the underamplified peak.  The Gaussian model is motivated by the work in \citeref{mayani}.

\subsection{\label{sec:level3-1}Hamamatsu R11410 Analysis}

The R11410 data includes a dedicated background measurement that can be used to constrain the model.  For example, an exponential contribution to the background (as suggested in \citeref{bellamy}) is ruled out and $\mu_{bkg}$ and $\sigma_{bkg}$ are constrained with a prior during the fits for both the cascade and Gaussian model (the background spectrum is shown figures 4, 6, and 7 in \citeref{saldanha}).

While in a standard experiment one could take multiple datasets while varying light levels and fit all data simultaneously to calibrate the SPE response, in this work it was decided to fit each light level individually in order to compare these results directly to the model independent method and the Gaussian model.  

The results of the fit are shown in \tabref{tab-1}.  In the table, $\eta$ is the attenuation factor of the filter in between the laser and PMT, $\lambda$ is the mean number of photoelectrons per light pulse, and $\mu$ and $\sigma$ are the mean and standard deviation of the resulting SPE response function.  All uncertainties shown are statistical.  Note that the model independent (MI) results and the cascade model (CM) results agree typically within a few percent and never disagree by more than 10\%.

Of all the parameters used in the cascade model and  described in \secref{sec:level2-4}, only $\mu_{epd}$ and $p_c$, which are primarily related to the mean of the individual dynode stages' responses, showed a high degree of correlation or anticorrelation.

\begin{table*}[b]
\centering
\caption{Comparison of model independent (MI), cascade model (CM), and Gaussian model (GM) using the R11410 PMT.}
\label{tab-1}
\def\arraystretch{1.2}
\resizebox{\textwidth}{!}{
\begin{tabular}{cc|ccccc}

\multicolumn{2}{c|}{Voltage [V]} & 1400 & 1500 & 1600 & 1700 & 1700 \\

\multicolumn{2}{c|}{$\eta$} & $2 \times 10^5$ & $2 \times 10^5$ & $2 \times 10^5$ & $2 \times 10^5$ & $1 \times 10^5$ \\ \hline

\multirow{3}{*}{$\lambda$} & MI & $1.257 \pm 0.005$ & $1.289 \pm 0.005$ & $1.324 \pm 0.005$ & $1.351 \pm 0.005$ & $2.395 \pm 0.008$ \\
						   & CM & $1.226^{+0.123}_{-0.043}$ & $1.256^{+0.010}_{-0.010}$ & $1.300^{+0.009}_{-0.008}$ & $1.315^{+0.008}_{-0.009}$ & $2.372^{+0.025}_{-0.022}$ \\
						   & GM & $1.188^{+0.022}_{-0.018}$ & $1.201^{+0.009}_{-0.008}$ & $1.234^{+0.011}_{-0.009}$ & $1.275^{+0.014}_{-0.013}$ & $2.284^{+0.029}_{-0.032}$ \\ \hline

$\lambda_{FA}$ & CM & $1.063^{+0.007}_{-0.007}$ & $1.039^{+0.007}_{-0.007}$ & $1.039^{+0.006}_{-0.006}$ & $1.042^{+0.006}_{-0.006}$ & $1.842^{+0.015}_{-0.016}$ \\ \hline

\multirow{3}{*}{$\mu$ [$\textrm{e}^-$]} & MI & $(1.88 \pm 0.01) \times 10^6$ & $(3.10 \pm 0.01) \times 10^6$ & $(4.98 \pm 0.02) \times 10^6$ & $(7.88 \pm 0.02) \times 10^6$ & $(7.90 \pm 0.02) \times 10^6$ \\
                       & CM & $(1.87 \pm 0.15) \times 10^6$ & $(3.17 \pm 0.02) \times 10^6$ & $(5.12 \pm 0.03) \times 10^6$ & $(8.03 \pm 0.04) \times 10^6$ & $(7.88 \pm 0.08) \times 10^6$ \\
                       & GM & $(1.96 \pm 0.03) \times 10^6$ & $(3.30 \pm 0.02) \times 10^6$ & $(5.34 \pm 0.05) \times 10^6$ & $(8.19 \pm 0.09) \times 10^6$ & $(8.08 \pm 0.07) \times 10^6$ \\ \hline
                       
\multirow{3}{*}{$\sigma$ [$\textrm{e}^-$]} 
					   & MI & $(8.6 \pm 0.1) \times 10^5$ & $(1.56 \pm 0.02) \times 10^6$ & $(2.84 \pm 0.02) \times 10^6$ & $(4.49 \pm 0.05) \times 10^6$ & $(4.51 \pm 0.05) \times 10^6$ \\
                       & CM & $(9.0 \pm 0.9) \times 10^5$ & $(1.56 \pm 0.01) \times 10^6$ & $(2.66 \pm 0.02) \times 10^6$ & $(4.26 \pm 0.03) \times 10^6$ & $(4.34 \pm 0.05) \times 10^6$ \\
                       & GM & $(7.7 \pm 0.3) \times 10^5$ & $(1.38 \pm 0.02) \times 10^6$ & $(2.35 \pm 0.02) \times 10^6$ & $(3.80 \pm 0.03) \times 10^6$ & $(3.81 \pm 0.03) \times 10^6$ \\ \hline
                       
\multicolumn{2}{c|}{$\textrm{ln} \left( \frac{\mathcal{L}_{CM}}{\mathcal{L}_{GM}} \right)$} & -14.5 & 17.2 & 257.4 & 567.2 & 183.8 \\

\end{tabular}
}
\end{table*}

Also of note in \tabref{tab-1} is the row $\lambda_{FA}$ denoting the mean number of photoelectrons fully amplified.  Unlike the other $\lambda$ measurements, $\lambda_{FA}$ should be approximately voltage independent since underamplification effects are removed.  As can be seen, all $\lambda_{FA}$ with the same attenuation $\eta$ agree within $\sim$1\%, providing a cross-check on the cascade model fit.

Another very important feature of the table is the last row, which compares the best-fit log-likelihood of the cascade model to the best-fit log-likelihood of the Gaussian model.  The cascade model significantly outperformed the Gaussian model in four out of five of the datasets.  Unsurprisingly, the dataset where the Gaussian model outperforms the cascade model is when the voltage is lowest and the valley in between the background and single photoelectron peak plays the smallest role in the fit.  This improvement is likely due to the increased freedom in the Gaussian model since the underamplfied peak is almost entirely independent of the fully-amplified peak.

\begin{figure}[t]
    \centering
    \begin{subfigure}[t]{0.49\textwidth}
        \centering
        \includegraphics[width=\linewidth]{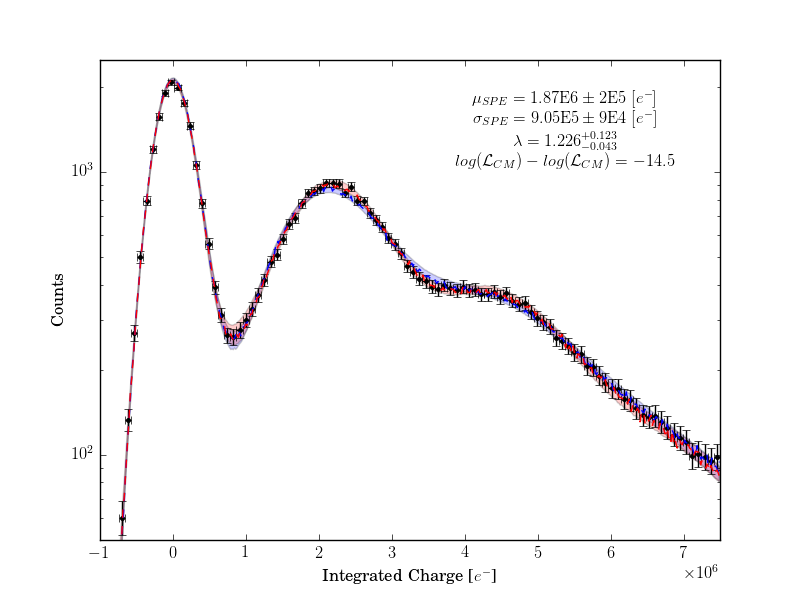} 
        \caption{R11410 PMT at 1400 V with attenuation of 2E5}
        \label{fig:67_68_best}
    \end{subfigure}
    \hfill
    \begin{subfigure}[t]{0.49\textwidth}
        \centering
        \includegraphics[width=\linewidth]{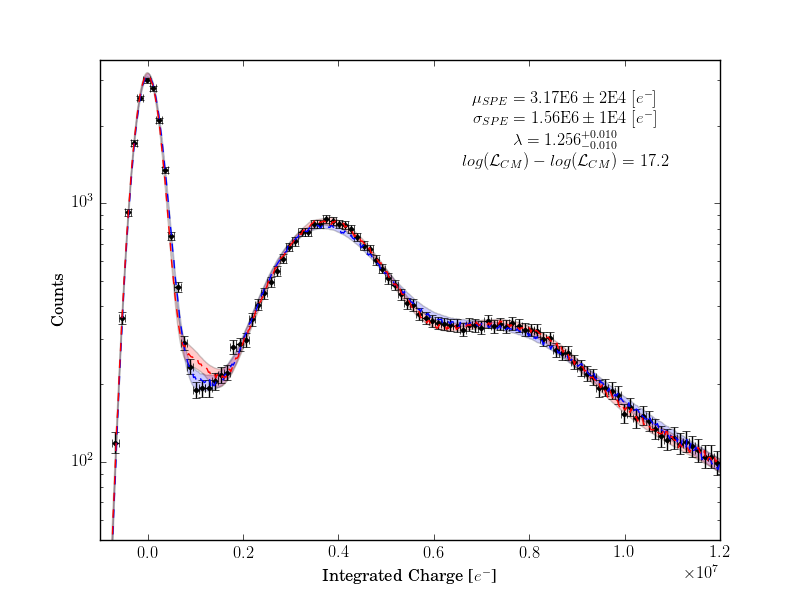} 
        \caption{R11410 PMT at 1500 V with attenuation of 2E5}
        \label{fig:66_65_best}
    \end{subfigure}

    
    \begin{subfigure}[t]{0.49\textwidth}
        \centering
        \includegraphics[width=\linewidth]{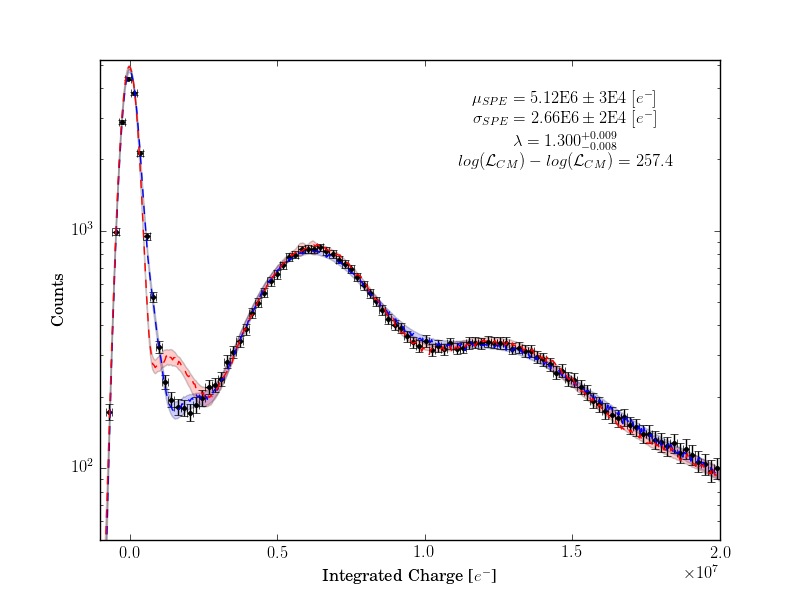} 
        \caption{R11410 PMT at 1600 V with attenuation of 2E5}
        \label{fig:62_61_best}
    \end{subfigure}
    \hfill
    \begin{subfigure}[t]{0.49\textwidth}
        \centering
        \includegraphics[width=\linewidth]{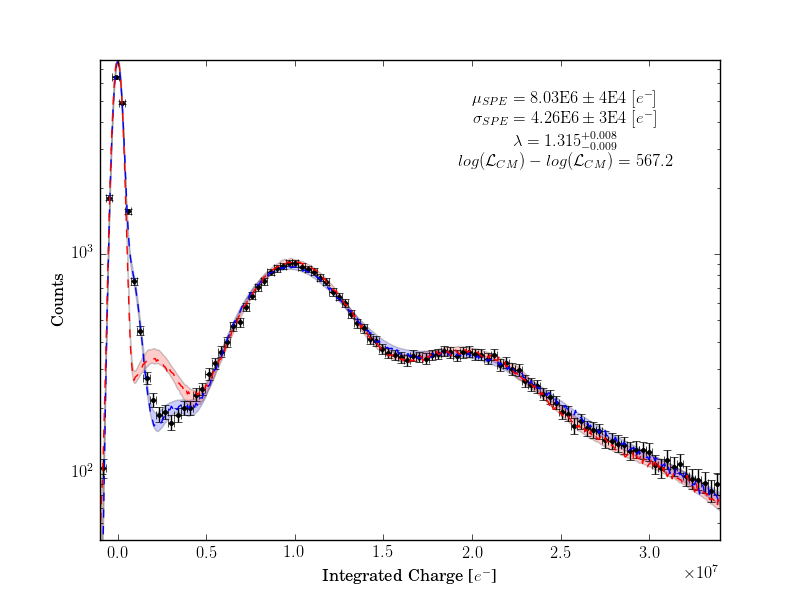} 
        \caption{R11410 PMT at 1700 V with attenuation of 2E5}
        \label{fig:72_71_best}
    \end{subfigure}


    \begin{subfigure}[t]{0.49\textwidth}
    \centering
        \includegraphics[width=\linewidth]{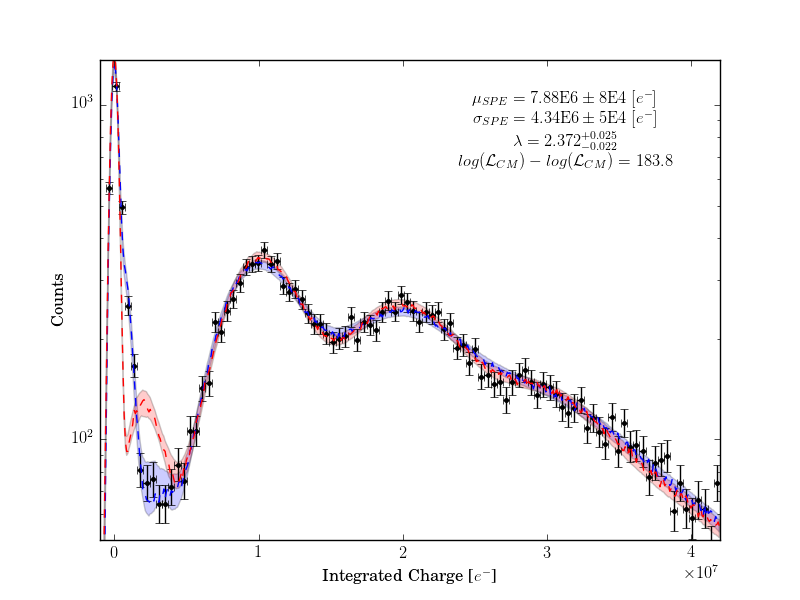} 
        \caption{R11410 PMT at 1700 V with attenuation of 1E5}
        \label{fig:73_74_best}
    \end{subfigure}
    \caption{The laser calibration charge spectra for the R11410 PMT at different voltages and attenuation levels with the best-fit models and 95\% confidence bands overlaid.  The cascade model is shown in blue while the Gaussian model is shown in red.  The statistics shown are for the cascade model.}
    
    \label{fig:uc_best_fits}
\end{figure}

\figref{fig:uc_best_fits} shows the best-fits for both the cascade model (blue) and the Gaussian model (red) compared to laser calibration charge data along with the 95\% confidence bands of each fit.  Notice that as the gain increases, the Gaussian model is unable to explain the behavior in the valley while the cascade model predicts this behavior well in all five spectra.



\begin{figure}[t]
\centering
\includegraphics[width=9cm]{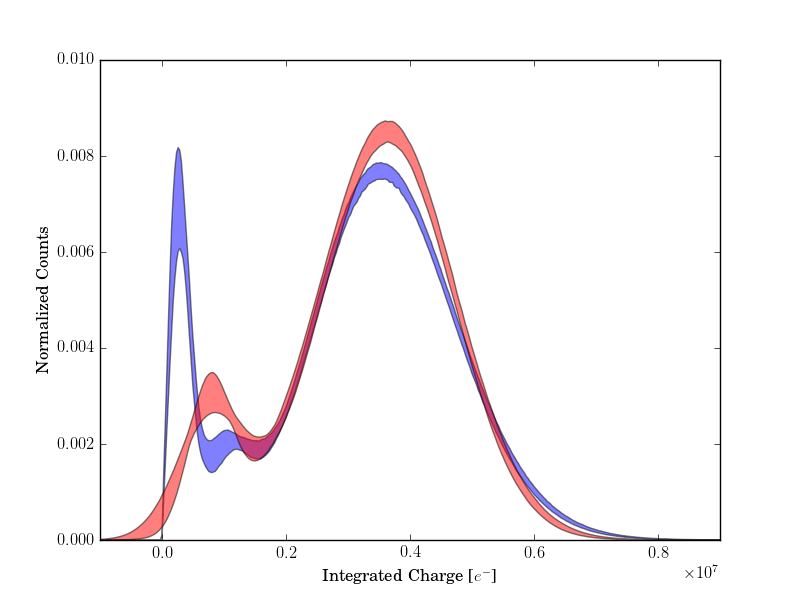}
\caption{The predicted SPE charge response for the R11410 PMT at 1500 V for the cascade model (blue) and the Gaussian model (red).  In the cascade model SPE charge response spectrum one can see, from left to right, the three major features: the underamplified peak from photons striking the first dynode, the underamplified peak from a non-ideal trajectory for electrons from the photocathode, and the fully-amplified peak.  At such a high gain the fully-amplified peak (right-most) of the cascade and Gaussian models agree.  However, there is a large discrepancy in the region of underamplified electrons.}
\label{fig:fig-2}
\end{figure}

In \figref{fig:fig-2} one can see the predicted SPE charge response for the R11410 PMT at 1500 V without background convolution.  The region shown is again the 95\% confidence interval.  One can see the three major features going from left to right: the underamplified peak from photons striking the first dynode, the underamplified peak from a non-ideal trajectory, and the fully-amplified peak.  Note that the signal can naturally never be less than zero unlike most analytical models that are either truncated or allowed to extend into a non-physical region.  The fully-amplified signal (right-most peak) is fairly symmetric - this is because this PMT operates at high gain and has good resolution.  This, however, will not be the case when looking at the fully-amplified peak for the R6041-406 PMT in \secref{sec:level3-2}.

\begin{table}
\centering
\caption{Goodness of Fit Tests For R11410 with Cascade Model.}
\label{tab-gof}
\begin{tabular}{|c|c|c|c|c|}
\hline
Voltage {[}V{]} & $\eta$ & Reduced $\chi^2$ & $p_{\chi^2}$ & $p_{KS}$  \\ \hline
1400 & $2 \times 10^5$ & 0.67 & 0.991 & 0.274 \\ \hline
1500 & $2 \times 10^5$ & 0.95 & 0.604 & 0.259 \\ \hline
1600 & $2 \times 10^5$ & 0.90 & 0.742 & 0.287 \\ \hline
1700 & $2 \times 10^5$ & 1.28 & 0.037 & 0.327 \\ \hline
1700 & $1 \times 10^5$ & 0.98 & 0.539 & 0.279 \\ \hline

\end{tabular}
\end{table}

Shown in \tabref{tab-gof} are the results of the goodness of fit tests for the best-fit parameters.  Since the parameter estimation is performed in a single dimension, one can look at the relatively simple $\chi^2$ test and the more robust Kolmogorov-Smirnov (KS) test.  While there is more fluctuation from the $\chi^2$ test, all tests show little or no evidence against the cascade model.

\subsection{\label{sec:level3-2}Hamamatsu R6041-406 Analysis}

\begin{table*}[t]
\centering
\caption{Comparison of cascade and Gaussian models using the R6041-406 PMT.}
\label{tab-nerix}

\def\arraystretch{1.2}
\begin{tabular}{cc|cc}

\multicolumn{2}{c|}{Voltage [V]} & 800 & 800 \\

\multicolumn{2}{c|}{Light Level} & \RNum{1} & \RNum{2} \\ \hline

\multirow{2}{*}{$\lambda$} & CM & $1.237^{+0.070}_{-0.042}$ & $2.455^{+0.148}_{-0.140}$ \\
						   & GM & $1.403^{+0.064}_{-0.046}$ & $2.709^{+0.108}_{-0.098}$ \\ \hline

\multirow{2}{*}{$\mu$ [$\textrm{e}^-$]} 
					   & CM & $(8.43 \pm 0.48) \times 10^5$ & $(8.53 \pm 0.68) \times 10^5$ \\
                       & GM & $(7.57 \pm 0.32) \times 10^5$ & $(7.50 \pm 0.23) \times 10^5$ \\ \hline
                       
\multirow{2}{*}{$\sigma$ [$\textrm{e}^-$]} 
						  & CM & $(5.61 \pm 0.17) \times 10^5$ & $(5.89 \pm 0.17) \times 10^5$ \\
                          & GM & $(5.79 \pm 0.11) \times 10^5$ & $(6.00 \pm 0.12) \times 10^5$ \\ \hline
                       
\multicolumn{2}{c|}{$\textrm{ln} \left( \frac{\mathcal{L}_{CM}}{\mathcal{L}_{GM}} \right)$} & 4.9 & 11.9 \\

\end{tabular}
\end{table*}

\begin{table}[h]
\centering
\caption{Goodness of Fit Tests For R6041-406 with Cascade Model.}
\label{tab-gof_nerix}
\begin{tabular}{|c|c|c|c|c|}
\hline
Voltage {[}V{]} & Light Level & Reduced $\chi^2$ & $p_{\chi^2}$ & $p_{KS}$  \\ \hline
800 & I & 1.39 & 0.009 & 0.361 \\ \hline
800 & II & 1.50 & 0.002 & 0.353 \\ \hline

\end{tabular}
\end{table}

In addition to the analysis performed with a PMT capable of large gains, the cascade model was also used to calibrate a PMT that must operate at significantly lower gains and with worse noise conditions.  While a background measurement was taken, since the calibration is done in situ with an LED and pulser it is impossible to confirm that noise conditions were the same between the dedicated background measurement (pulser off) and the measurements with the pulser on.  For this reason and given that the width of the background peak is on the order of the SPE response, the model independent approach cannot be used.

Since the dedicated background measurements for this PMT could not be used, one would normally try multiple background models to study the potential systematic effects.  However, in this work, only the Gaussian background model is examined for consistency.

In this specific calibration, two light levels were used which are denoted \RNum{1} and \RNum{2} corresponding to different pulser voltages used in conjunction with a blue LED.  While for the detector discussed in \citeref{goetzke} these two light levels were fit simultaneously, only the results from individual fits are shown for consistency.

\begin{figure}[t]
    \centering
    \begin{subfigure}[t]{0.49\textwidth}
        \centering
        \includegraphics[width=\linewidth]{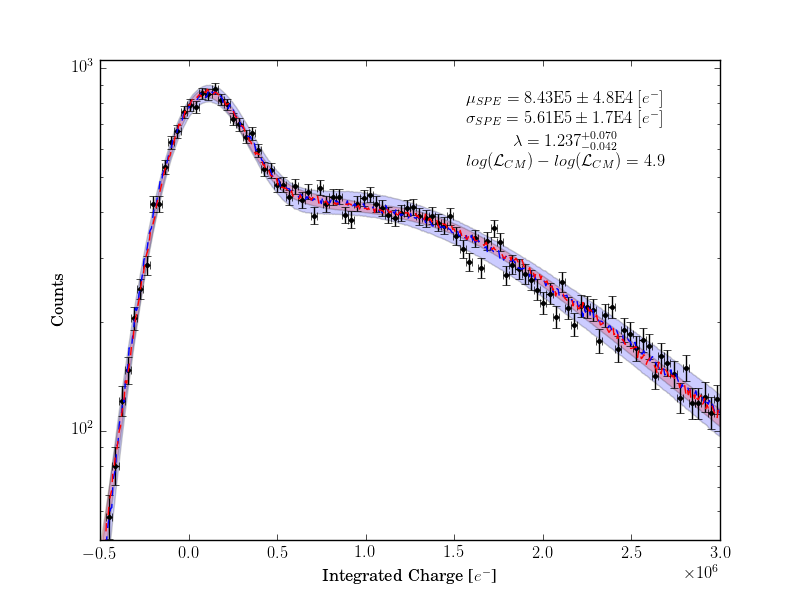} 
        \caption{R6041-406 PMT at 800 V at light level I}
        \label{fig:1523_best}
    \end{subfigure}
    \hfill
    \begin{subfigure}[t]{0.49\textwidth}
        \centering
        \includegraphics[width=\linewidth]{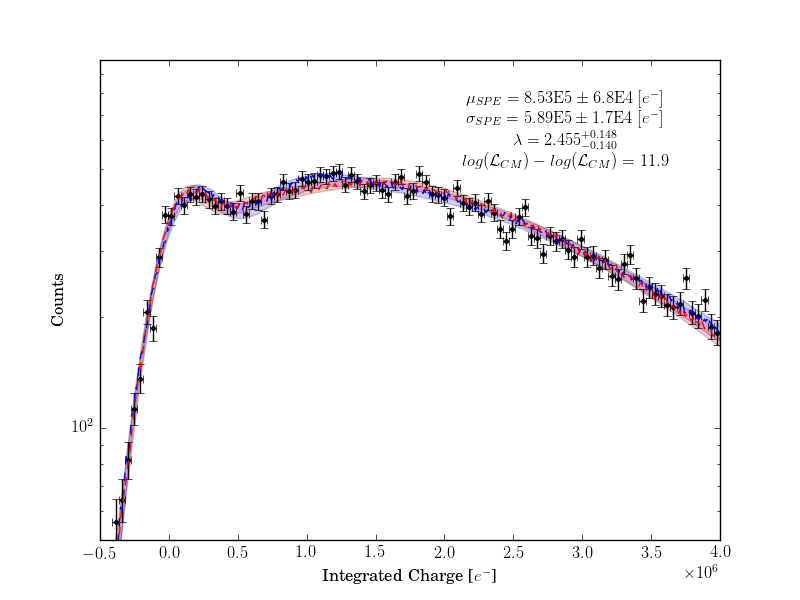} 
        \caption{R6041-406 PMT at 800 V at light level II}
        \label{fig:1531_best}
    \end{subfigure}

    \caption{The diode calibration charge spectra for the R6041-406 PMT at 800 V with the best-fit models and 95\% confidence bands overlaid.  The cascade model is shown in blue while the Gaussian model is shown in red.  The statistics shown are for the cascade model.}
    
    \label{fig:nerix_best_fits}
\end{figure}

The results of the parameter estimation and the goodness of fit tests are shown in \tabref{tab-nerix} and \tabref{tab-gof_nerix}.  The best fit and the 95\% confidence band for each light level can be found in \figref{fig:nerix_best_fits}.  While the $\chi^2$ test shows evidence against the cascade model, it seems that this is solely due to the behavior in a handful bins that fall outside of the 95\% confidence band as seen in both spectra in \figref{fig:nerix_best_fits}.  This hypothesis is further supported by the results of the Kolmogorov-Smirnov test, which shows no evidence against the cascade model.  As a further cross-check, one can also compare the mean and standard deviation of the response function from both light levels which agree with each other well within uncertainty.

As with the Hamamatsu R11410, of all the parameters used in the cascade model and  described in \secref{sec:level2-4}, only $\mu_{epd}$ and $p_c$, which are primarily related to the mean of the individual dynode stages' responses, showed a high degree of correlation or anticorrelation.

\begin{figure}[h]
\centering
\includegraphics[width=9cm]{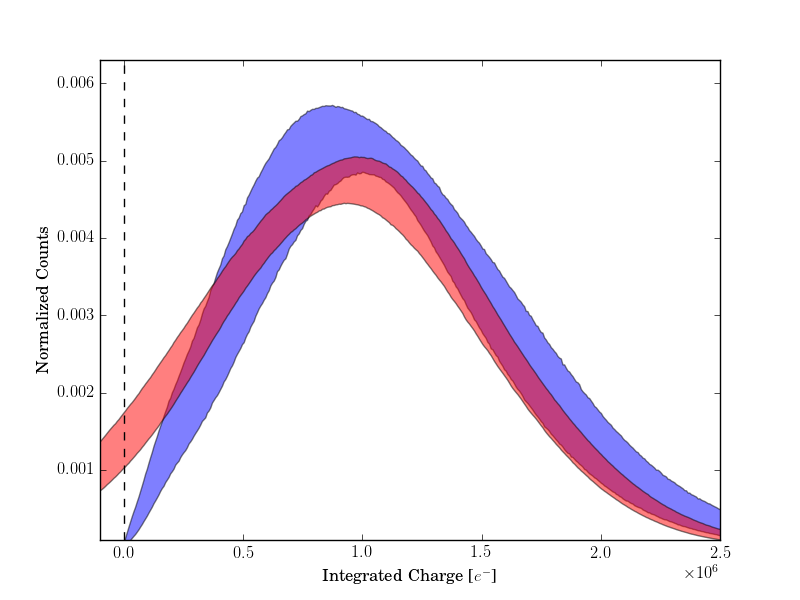}
\caption{The predicted fully-amplified photoelectron response for the R6041-406 PMT at 800 V.  Notice the asymmetry in the cascade model response (shown in blue) and how far into the unphysical regime the Gaussian model response goes (shown in red). }
\label{fig:fig-nerix_spe}
\end{figure}

While the cascade model outperforms the Gaussian model in fit quality, as seen in the log-likelihood difference, the real power of the cascade model can be seen in \figref{fig:fig-nerix_spe}, which shows the fully-amplified peak only for both models without background convolution.  Notice again that the cascade model naturally begins at zero signal and has the asymmetry that one would expect while the Gaussian model predicts negative signal roughly 15\% of the time from the fully-amplified peak.  Clearly this prediction is not physical and would cause issues in MC simulations of the PMT.

\section{\label{sec:level4}Conclusions}

While the form of the cascade model presented in this paper will change for each type of PMT used in a different setting, we have shown that for the PMTs used in these two experiments that the cascade model is a much more realistic approximation of the photomultiplication process, agrees well with data, and is a drastic improvement in almost all cases versus the Gaussian model that is typically used for characterization of photomultipliers.  It is important that in future applications, the analyzer checks different sources of underamplified electrons and different background models if a dedicated background measurement was not performed.  Parameters of the model may be further constrained by estimating them with multiple light levels fit simultaneously.

It is our recommendation that the cascade model be used in conjunction with the model independent prescription described in detail in \citeref{saldanha}.  Since it is relatively unlikely for a PMT's characteristics to change during a measurement, we recommend that an initial characterization be performed using the cascade model and cross-checked with the model independent estimation.  Following the initial characterization, performance can be monitored solely by the model independent estimation with the cascade model reserved for spot checks and diagnosis if PMT performance changes.

\acknowledgments

We gratefully acknowledge support from the National Science Foundation at Columbia University (Grant No. PHYS09-04220) and at the University of Chicago (Grant No. PHY-1505581).


\begin{thebibliography}{24}

\bibitem{bellamy} 
E. Bellamy, et al., \textit{Absolute calibration and monitoring of a spectrometric channel using a photomultiplier}, Nucl. Instrum. Methods Phys. Res. A, vol. 339, pp. 468--476 (1993).

\bibitem{dossi} 
R. Dossi, et al., \textit{Methods for precise
photoelectron counting with photomultipliers}, Nucl. Instrum. Methods Phys. Res. A, vol. 451, pg. 623--637 (2000).

\bibitem{haas} J. T. M. de Haas and P. Dorenbos, \textit{Methods for accurate measurement of the response of photomultiplier tubes and intensity of light pulses}, Nuclear Science, IEEE Transactions, pg. 205--210 (2011).

\bibitem{saldanha}
R. Saldanha, L. Grandi, Y. Guardincerri and T. Wester.  \textit{Model Independent Approach to the Single Photoelectron Calibration of Photomultiplier Tubes}, Nucl. Instrum. Methods Phys. Res. A, vol. 863, pg. 35--46 (2017).

\bibitem{lombard}
F.J. Lombard, F. Martin. \textit{Statistics of electron multiplication}, Rev. Sci. Instrum. 32, pg. 200–-201 (1961).

\bibitem{prescott}
J. Prescott, \textit{A statistical model for photomultiplier single-electron statistics}, Nucl. Instrum. Methods 39, pg. 173-–179 (1966).



\bibitem{hammamatsu}
D. Carter, \textit{Photomultiplier Handbook: Theory, Design, Application}, Lancaster, Pennsylvania: Burle Industries, Inc. (1980).

\bibitem{dpe}
C. H. Faham, V. M. Gehman, A. Currie, A. Dobi, P.
Sorensen, and R. J. Gaitskell, \textit{Measurements of wavelength-dependent
double photoelectron emission from
single photons in VUV-sensitive photomultiplier tubes},
J. Instrum. 10, P09010 (2015).

\bibitem{tan}
H. Tan. \textit{A statistical model of the photomultiplier gain process with applications to optical pulse detection}, Proceedings of the ITC, pg. 115 (1982).

\bibitem{dwarakanath}
N.C. Dwarakanath,S.D. Galbraith. \textit{Sampling from discrete Gaussians for lattice-based cryptography on a constrained device}, Applicable Algebra in Engineering, Communication and Computing, pg. 1--22 (2014).

\bibitem{dfm}
Daniel Foreman-Mackey, David W. Hogg, Dustin Lang and Jonathan Goodman.
\newblock \textit{emcee: The MCMC Hammer} (2012);
\newblock arXiv:1202.3665.
\newblock DOI: 10.1086/670067.

\bibitem{xenon1t}
XENON Collaboration (E. Aprile et al), \textit{Physics reach of the XENON1T dark matter experiment}, JCAP 1604 (2016); arXiv:1512.07501.
 
\bibitem{goetzke}
L. W. Goetzke, E. Aprile, M. Anthony, G. Plante and M. Weber.  \textit{Measurement of light and charge yield of low-energy electronic recoils in liquid xenon} (2016);
arXiv:1611.10322.


\bibitem{mayani}
D. Mayani. \textit{Photomultiplier Tubes for the XENON1T Dark Matter Experiment and Studies on the XENON100 Electromagnetic Background}, PhD Thesis (2017), University of Zurich.

\end{thebibliography}
\end{document}